\documentclass[aps,pra,twocolumn,groupedaddress,showpacs]{revtex4-1}

\usepackage{dcolumn}

\begin{document}

%\preprint{}

\title{Experimental isotope shifts of the 5 $^2S_{1/2}$ state and low-lying excited states of Rb}


\author{L. Aldridge}
\author{P. L. Gould}
\author{E. E. Eyler}
%\thanks{}
\affiliation{Physics Department, University of Connecticut, Storrs, CT 06269}


\date{\today}

\begin{abstract}
By combining a recent precise measurement of the ionization energy of $^{87}$Rb with previous measurements of electronic and hyperfine structure, an accurate value for the $^{85}\textrm{Rb}-^{87}\textrm{Rb}$ isotope shift of the 5$^2S_{1/2}$ ground state can be determined.  In turn, comparison with additional spectroscopic data makes it possible for the first time to evaluate isotope shifts for the low-lying excited states, accurate in most cases to about 1 MHz.  In a few cases, the specific mass shift contribution can be determined in addition to the total shift.  This information is particularly useful for spectroscopic analysis of transitions to Rydberg states, and for tests of atomic theory.
\end{abstract}

\pacs{32.30.Jc,31.30.Gs,32.80.Ee}
%\keywords{}

\maketitle


 Much experimental work has been done on measuring with high accuracy the absolute frequencies of various ground state to singly-excited state transitions in the two most prevalent rubidium isotopes, $^{85}$Rb and $^{87}$Rb \cite{Barwood1991,Chui2005,Sansonetti2006,Nez1993,Moon2009,Banerjee2003,Grundevik1977}.   In many cases transition frequencies have been measured for both isotopes, allowing accurate determinations of the transition isotope shift.  However, isotope shifts for individual electronic states of rubidium have not been previously reported, with the exception of the ground 5 $^2S_{1/2}$ state.  Even for the ground state, the best previous determination is accurate only to 10 MHz \cite{Stoicheff1979}.  Accurate state-by-state isotope shifts are useful both for testing atomic many-body theory and for analyzing spectroscopic transitions between excited states in mixed-isotope samples.  This is especially useful for transitions to Rydberg states, which have negligible isotope shifts.  For example, this information can enable spectroscopic detection of near-resonant excitation exchange or charge exchange between ultracold isotopes of the same species.  The analysis of isotope shifts is also one of the most sensitive methods for determining isotopic variations in nuclear charge distributions \cite{Thibault1981,Angeli2004}.

 Isotope shifts arise due to three distinct effects.  The normal mass shift (NMS) is due to the differing isotopic mass, and thus differing reduced mass in the kinetic energy term of the Hamiltonian.  The specific mass shift (SMS) is dependent on electron-electron correlations as well as the reduced mass, and the field shift (FS) is the contribution due to the spatial distribution of the nuclear charge density \cite{Hartley1991,Johnson2007}.  Generally speaking, the SMS poses difficult challenges for atomic theory \cite{Safronova2005}, while the FS is more tractable, and is closely linked to the mean-squared nuclear charge radius. In the specific cases of $^{87}$Rb and $^{85}$Rb, the charge radii are well-known from muonic X-ray data \cite{Fricke1995,Angeli2004,Orozco2008}, and experimental isotope shifts can be regarded as benchmarks for theoretical calculations of the SMS \cite{Safronova2005}.

 Major improvements in the experimental isotope shift data for several electronic states of rubidium can be realized by combining an impressively accurate new measurement of the ionization energy (IE) of $^{87}$Rb \cite{Mack2011} with previous spectroscopic data for electronic and hyperfine structure.  The new $^{87}$Rb results are accurate to 0.3 MHz, an improvement over previous work \cite{Stoicheff1979} by more than two orders of magnitude.  To determine the $5\,^2S_{1/2}$ ground state isotope shift, the $^{87}$Rb IE can be combined with the $^{85}$Rb IE determined from earlier high-precision measurements of Rydberg $ns$ states \cite{Sansonetti1985,Sansonetti2006}.  We take the accuracy of the $^{85}$Rb IE to be 0.9 MHz as listed in the recent review article by J. Sansonetti (Ref. \cite{Sansonetti2006}), but we caution that the two data fitting methods used in the original experimental analysis differ by 2.1 MHz \cite{Sansonetti1985}, suggesting that the uncertainty may be slightly underestimated.

To remove the hyperfine structure from the experimental results, the position of the hyperfine center-of-mass (COM) energy must be determined.  A conventional two-parameter model \cite{Arimondo1977} is sufficient to describe the hyperfine structure to the required accuracy,
 \begin{equation}
E_i(F) = \frac{1}{2} A_i K + B_i \frac{\frac{3}{2} K (K+1) - 2 I (I+1) J (J+1)}{2I (2I-1) 2J (2J-1)}.
\end{equation}
Here $K = F(F+1) - I(I+1) - J(J+1)$, and $A_i$ and $B_i$ are the magnetic dipole and electric quadrupole hyperfine constants for state $| i \rangle$, with $B_i$ contributing only when $J > 1/2$.

 \begin{table*}
 \caption{\label{ShiftTable}Isotope shifts $\delta^{85-87}$ for selected low-lying states of rubidium\emph{}.}
 \begin{ruledtabular}
 \begin{tabular}{l d d d l}
 State & \multicolumn{1}{c}{Total Isotope} & \multicolumn{1}{c}{Normal Mass} & \multicolumn{1}{c}{Specific Mass +} & Refs\\
  & \multicolumn{1}{c}{Shift (MHz)} & \multicolumn{1}{c}{Shift (MHz)} & \multicolumn{1}{c}{Field Shifts (MHz)} & \\
  \hline
$5\,^2S_{1/2}$ & 164.35\,(95)\footnote{Uncertainties may be slightly larger due to a possible underestimation in Ref. \cite{Sansonetti2006}; see text.} & 149.97 & 14.38\,(95) & \cite{Sansonetti2006,Mack2011,Bize1999} \\
$7\,^2S_{1/2}$ & 32.79\,(95) & 32.85 & -0.06\,(95) & \cite{Chui2005} \\
\\
$5\,^2P_{1/2}$ & 86.77\,(95)  & 93.98  & -7.21\,(95) & \cite{Barwood1991}\\
$5\,^2P_{3/2}$ & 86.31\,(95) & 92.92 & -6.61\,(95) & \cite{Barwood1991,Banerjee2003}\\
$6\,^2P_{1/2}$ & 40.1\,(10) & 44.4 & -4.3\,(10) & \cite{Grundevik1977} \\
$6\,^2P_{3/2}$ & 40.2\,(12) & 44.1 & -3.9\,(12) & \cite{Grundevik1977} \\
\\
$4\,^2D_{3/2}$ & 1.8(17) & 63.8 & -62.0\,(17) & \cite{Moon2009,Arimondo1977}\\
$5\,^2D_{3/2}$ & -0.88\,(95) & 35.57 & -36.45\,(95) & \cite{Nez1993} \\
$5\,^2D_{5/2}$ & 1.32\,(95) & 35.55 & -34.23\,(95) & \cite{Nez1993} \\

 \end{tabular}
 \end{ruledtabular}
 \end{table*}

  For the $5\,^2S_{1/2}$ ground state of $^{87}$Rb, the IE reported in Ref. \cite{Mack2011} is referenced to the $F$=1 hyperfine level of the ground state, and we used accurate measurements of the hyperfine structure from Refs. \cite{Nez1993} and \cite{Bize1999} to refer the IE to the COM prior to comparison with $^{85}$Rb. The resulting ground-state isotope shift is $\delta^{85-87}= 164.35 \pm 0.95$ MHz, where $\delta^{85-87} \equiv E_{85} - E_{87}$.  This ground-state isotope shift is consistent with the best previous determination of $167 \pm 10$ MHz \cite{Stoicheff1979}, but is an order of magnitude more accurate.  Given this result, isotope shifts for several excited electronic states can in turn be calculated by use of existing high-precision measurements of electronic transitions and hyperfine structure.  In Table \ref{ShiftTable} we collect the results.  In most cases the original data analyses included calculations of transition isotope shifts referenced to the hyperfine COM, and it was necessary only to subtract out the ground-state isotope shift. The uncertainty listed for the total isotope shift of each state is the quadrature sum of the uncertainty in the ground-state isotope shift and the the transition isotope shift for that state. Generally the ground-state isotope shift dominates the uncertainty.

  In the case of the $4\,^{2}D_{3/2}$ state, the transition isotope shift was not calculated in the most recent previous work \cite{Moon2009} and only two transitions per isotope were measured, not enough to determine the hyperfine constants without further information. In that work, the $A$ and $B$ constants were determined by assuming that the ratio $A_{87}/A_{85}$ is the same as for the $5\,^{2}S_{1/2}$ state, and that the ratio $B_{87}/B_{85}$ is the same as for the $5\,^{2}P_{3/2}$ state. However, examination of hyperfine data for several electronic states shows that the typical hyperfine anomalies (i.e., deviations from this assumption) are large enough to affect the $A$ and $B$ constants by several percent. An interesting discussion of some of the origins of hyperfine anomalies for rubidium is provided in Ref. \cite{Orozco2008}.  We chose in this work to use previous experimental values for $A$ \cite{Arimondo1977} that did not rely on this scaling argument.  However, we retained the values for the quadrupole hyperfine constant $B$ from Ref. \cite{Moon2009} as no other determinations are available. Fortunately the $B$ constants are small enough ($B_{87}$=2.2 MHz and $B_{85}$=4.5 MHz) that a variation by several percent would not significantly affect our results. Another special case is the $5\,^{2}P_{3/2}$ state, for which two disagreeing measurements of the transition isotope shift exist, with approximately equal quoted accuracies\cite{Barwood1991,Banerjee2003}. The difference of 0.41 MHz is significantly smaller than our overall uncertainty of about 1 MHz, and we chose in this work simply to use an average of the two values.

  In Table I we also show the normal mass shift, easily calculated using the COM energies for the $\,^{85}$Rb levels as extracted from Refs. \cite{Barwood1991,Chui2005,Sansonetti2006,Nez1993,Moon2009,Banerjee2003,Grundevik1977} and isotopic masses from Ref. \cite{Audi2003}, and the remaining contribution due to the combined SMS and FS.  The results for $nd$ states reveal a striking accidental cancellation that leaves a net isotope shift of very nearly zero, whereas for the $ns$ and $np$ states the normal mass shift is dominant.  It would be of considerable interest to further separate the FS from the SMS contributions, which would provide experimental benchmarks for the SMS.  In principle the field shift can be used to determine the isotopic difference in the nuclear mean-squared charge radius, but as already mentioned these radii are already known to about 0.04\% \cite{Angeli2004,Orozco2008}.   For $^{87}$Rb there is an interesting minimum in $\langle r^2 \rangle$ because the neutron number $A = 50$ is a ``magic'' value that leads to closed-shell nuclear configuration \cite{Maharana1996}.  As a result there have been numerous analyses and calculations of Rb isotope shifts as a function of $N$, including recent nuclear structure calculations that are beginning to approach experimental accuracy \cite {Rodriguez2010}.

 The FS and SMS terms could easily be separated if the FS is assumed to be dominated by a simple contact interaction, but calculations for Na, K, Cs, and Fr \cite{Hartley1991,Safronova2001,Safronova2005} show that higher-order contributions can be significant, and there are small FS contributions even for states with with $\ell\neq 0$.  Thus accurate calculations are needed.  For Rb these field-shift constants have been calculated only for the $5\,^2S_{1/2}$ and $5\,^2P_{3/2}$ states \cite{Safronova2005}.  Using $\delta\langle r^2 \rangle = 0.042$~fm$^2$ \cite{Angeli2004,Orozco2008}, the calculated field shifts are 23.2 MHz for $5\,^2S_{1/2}$ and -0.66 MHz for $5\,^2P_{3/2}$.  Using Table I, the corresponding values for the SMS are -8.8 MHz and -6.0 MHz, respectively.  At present, reliable theoretical calculations of the SMS do not exist for Rb because of the challenge posed by near-cancellations of the various contributions \cite{Safronova2005,Safronova2011}, but for Na, K, Cs, and Fr calculations show substantial SMS contributions that vary almost unpredictably.  For example, they happen to be particularly large for the $3d$ levels of K, where the SMS is 1.08 times as large as the NMS, and has the opposite sign.  We hope that the present experimental isotope shifts will stimulate new theoretical calculation for Rb.

 In conclusion, we have determined isotope shifts $\delta^{85-87}$ accurate to about 1 MHz for the ground state and seven excited states of rubidium.  The ground-state result is consistent with prior determinations, while the excited-state shifts have not been previously determined.  For two states, $5\,^2S_{1/2}$ and $5\,^2P_{3/2}$, the specific mass shift could be determined as well.  These results should prove useful both for the analysis of excited-state spectra and for tests of atomic theory.

\begin{acknowledgments}
This work was supported by the National Science Foundation.  We would like to thank Dr. David Rahmlow and Prof. Marianna Safronova for clarifying discussions and helpful information.
\end{acknowledgments}

% Reference section:

\end{document}